\documentclass[reprint,twocolumn,aps,epsf,pre]{revtex4}
\usepackage{graphicx}
\usepackage{amsmath}
\usepackage{amssymb}
\usepackage{amsfonts}
\newtheorem{theorem}{Theorem}

\begin{document}
\title{Network Transfer Entropy and Metric Space for Causality Inference}
\author{Christopher R. S. Banerji$^{1,2,3}$, Simone Severini$^{1,4}$, Andrew E. Teschendorff$^{3}$\\
  \small $^1${Department of Computer Science, University College London, London WC1E 6BT, UK.}\\
  \small $^2${Centre of Mathematics and Physics in the Life Sciences and Experimental Biology, University College London, London WC1E 6BT, UK.}\\
  \small $^3${Statistical Cancer Genomics, Paul O'Gorman Building, UCL Cancer Institute, University College London, London WC1E 6BT, UK.}\\
  \small $^4${Department of Physics and Astronomy, University College London, London WC1E 6BT, UK.}
  \small e-mail address: \emph{christopher.banerji.11@ucl.ac.uk}}
\begin{abstract}
A measure is derived to quantify directed information transfer between pairs of vertices in a weighted network, over paths of a specified maximal length. Our approach employs a general, probabilistic model of network traffic, from which the informational distance between dynamics on two weighted networks can be naturally expressed as a Jensen Shannon Divergence. Our \emph{network transfer entropy} measure is shown to be able to distinguish and quantify causal relationships between network elements, in applications to simple synthetic networks and a biological signalling network. We conclude with a theoretical extension of our framework, in which the square root of the Jensen Shannon Divergence induces a metric on the space of dynamics on weighted networks. We prove a convergence criterion, demonstrating that a form of convergence in the structure of weighted networks in a family of matrix metric spaces implies convergence of their dynamics with respect to the square root Jensen Shannon Divergence metric.\end{abstract}
\maketitle
\section{Introduction}
Complex systems in diverse fields are often represented as weighted networks \cite{multi, wtw, airport}. Inferring information transfer between network elements from such representations can provide important insights into system structure and perturbations \cite{Kim, social,tan}. This paper proposes a general methodology to quantify such transfer based on information theory and probability on graphs.\\
Previous attempts to infer dynamics from weighted networks include an interaction model based on electrical circuitry, to discover active pathways contributing to the pathogenesis of the brain cancer Glioblastoma multiforme \cite{Kim,tan}. Reference \cite{social} uses a model of infection transmission, proportional to interaction frequency, to identify the spread of disease through social networks. Such case by case approaches have proved informative, however they are often tailor made to their applications and the {\it general} quantification of information transfer in weighted networks currently lacks a theoretical foundation.\\
The purpose of this paper is to construct an information theoretic measure, {\it network transfer entropy}, quantifying the directed amount of information transferred between any two vertices in a weighted network, with minimal assumptions and general applicability. Following construction (Section II), we demonstrate the measure on simple synthetic networks and a biological signalling network (Section III).\\
In the construction of general measures, one aims at an insight into the theoretical concepts governing the process being studied. We demonstrate that the network transfer entropy framework can be interpreted in the context of metric spaces (Section IV). In this construction one defines a family of mappings from the space of weighted networks (represented by matrices) to a family of metric spaces, whose elements describe possible signal dynamics on networks.\\ We prove a convergence principle, demonstrating that a form of convergence of weighted networks in the metric space $L^{p}(\mathbb {M}^{N\times N})$ implies convergence in the constructed metric space of signal dynamics. This result shows that in our general framework, deformation of network structure influences network dynamics in an intuitive way. This result has real world implications in, for example, network drug design, where one wishes to modify the chemical affinities of interacting proteins in a pathological signalling network ({\it i.e. } modify the edge weights) to restore a healthy signalling regime ({\it i.e.} modify the dynamics). Finally, we motivate how certain further theoretical problems in network evolution and network perturbation can be approached within this framework.\\
\section{Network Transfer Entropy}
{\it Transfer entropy} was introduced by Schreiber, to quantify the directed amount of information transferred between two mutually dependent time series \cite{shri}. This problem shares several important qualities with our problem of information transfer between network vertices, thus we follow Schreiber's approach in the derivation of our measure.\\
The definition of transfer entropy required a model in which it was possible to express whether two time series influenced each other. For transfer entropy to be widely applicable, this model needed to be sufficiently general to portray a wide array of diverse systems. It was thus intuitive to describe time series as realisations of (approximately) Markov processes of order $k$. For such a process $I$ the conditional probability of finding the process in state $i_{n+1}$ at time $n+1$ satisfies
\begin{eqnarray}
p(i_{n+1}|i_n,\ldots ,i_{n-k+1})=p(i_{n+1}|i_n,\ldots ,i_{n-k}).
\end{eqnarray}
These generalised Markov process are not all encompassing in their descriptive power; for example, they are in general not-applicable to studying subsystems of Markov processes \cite{vanKampen}. However for a broad range of datasets including heart and breathing rate data \cite{shri}, magnetoencephalography data \cite{mag} and financial time series \cite{finte}, the approximate Markov process model can be justified, making transfer entropy widely applicable.\\
The choice of a general dynamic model for a weighted network requires consideration of the literature. One must be careful to ensure the model makes minimal assumptions yet has sufficient descriptive power to portray complex systems. Much work has focused on interaction models known as {\it flow networks} (see for example \cite{flow}), in which transport from source nodes to sink nodes is subject to edge weight dependant constraints. These models are useful in optimisation problems where one wants to find paths through a network that maximise or minimise some function associated with path traversal, and thus tend to be used in systems where traffic can be manipulated, such as supply management \cite{supp_man}.\\
Flow networks are less useful in the interrogation of network dynamics where constraints on traffic are unknown, and sink and source nodes and not readily defined. Moreover, when network dynamics are stochastic and bursty rather than continuous flows, such as in social communication systems \cite{burstsoc} and gene regulatory networks \cite{burstgen}, adaptations of flow networks are required. Such adaptations include discrete flow networks \cite{dfn} and stochastic flow networks \cite{sfn}, in which the interaction of a vertex with neighbours is given by a probability distribution proportional to the edge weight distribution. The elegance of these discrete models is that they may approximate continuous models (such as flow networks) in the large time limit. Such models are generalised, for example by the inclusion of holding rates, in queuing theory \cite{SN} which with detailed information for parameter estimation can be used to describe and simulate a large variety of real world systems.\\
Given this literature we take our dynamic model for weighted networks as a balance between the descriptive power of stochastic networks of queuing theory and the simplicity of the stochastic flow networks. We consider the following Markovian model for signal dynamic evolution. Each vertex is assigned a data derived value quantifying a signal that the given vertex is capable of forwarding to one its neighbours. The vector containing these values for every vertex is referred to as the \emph{initial signal distribution} (ISD) of the network. In real world applications this distribution can be qualitatively diverse. For example, in biological networks, where vertices represent genes, the ISD may quantify the differential expression of genes in pathological versus healthy samples. In the airport transportation network where vertices are airports and edges connect airports one can fly between directly, the ISD may be the number of flights departing from each airport over a given time frame \cite{airport}. There are no restrictions on the ISD other than it being a vector in $\mathbb{R}^{N}$, where $N$ is the number of vertices.\\
We evolve this signal over the network $W=(w_{ij})_{i,j=1}^{N}$, where $w_{ij}>0$ for $i,j=1,...,N$, via a stochastic matrix $P=(p_{ij})_{i,j=1}^{N}$ defined as
\begin{eqnarray}
p_{ij}= \begin{cases}
   \frac{w_{ij}}{\sum_{j \in \mathcal{N}_{i}} w_{ij}} & \text{if $\sum_{j \in \mathcal{N}_{i}} w_{ij} \not= 0$} \\
   \delta_{i}^{j}       & \text{otherwise}
  \end{cases}
\end{eqnarray}
where $\mathcal{N}_{i}$ denotes the set of neighbours of vertex $i$ and $\delta_{i}^{j}$ denotes the Kronecker delta of $i$ and $j$. We evolve the ISD over multiple discrete time steps. At each time step the signal at each vertex $i$ is independently forwarded to vertex $j \in \mathcal{N}_{i}$ with probability $p_{ij}$ (we emphasise that self-edges can be added to the network and the weights on such edges would determine the probability that a vertex maintains its signal over a single time step). Thus the number of time steps directly corresponds to the maximal path length the ISD has traversed.\\
Given an ISD, $\vec{ X}_{0}$, and path length, $n$, for every vertex $i$ in the network we can compute the probability distribution of the signal at vertex $i$ given that the ISD has been forwarded along paths of length $n$ (see Appendix B). We denote this distribution $P[X^{i}_{n}|\vec{ X}_{0}]$.\\
Given a model of interactions, we wish to quantify paths of high or low traffic through the network. To proceed in this aim it suffices to identify the directed amount of information transferred between any two pairs of network vertices during a period of system evolution. To achieve this we again turn to Schreiber's methodology. In the derivation of transfer entropy the directed amount of information transferred from a process $J$ to a process $I$ is formulated as the incorrectness of the assumption that $I$ is not conditional on $J$. This quantity can be expressed as the {\it Kullback-Leibler divergence} \cite{Cover_thomas}
\begin{eqnarray}
\sum p(i^{n+1}_{n-k+1},j^n_l)\log \frac{ p(i_{n+1}|i^n_{n-k+1},j^n_l)}{p(i_{n+1}|i^n_{n-k+1})},
\end{eqnarray}
where $i_m^{n} = (i_n,\ldots ,i_{m})$ for $m<n$. Thus to quantify the amount of information vertex $j$ in our network transfers to vertex $i$ over paths of length $n$ we must derive a distribution for $X^{i}_{n}$ in which vertex $j$ sends no information to vertex $i$. We must then compute the informational distance between this distribution and the above distribution $P[X^{i}_{n}|\vec{{X}}_{0}]$ in which $j$ is able to communicate with $i$. Clearly if we set the $j^{th}$ row in the stochastic matrix $P$ to $\vec{ e}_j$ ({\it i.e.,} make $j$ an absorbing state; $\vec{ e}_{j}$ denotes the $j^{th}$ element of the standard basis of $\mathbb{R}^{N}$), then it is impossible for vertex $j$ to communicate with any vertex $i \not = j$ under our model. Given this modified matrix we can compute the probability distribution of $X^{i}_{n}$ given the ISD and that $j$ cannot communicate with $i$. We denote this distribution $P[X^{i}_{n}|\vec{{X}}_{0},j]$. Here we diverge from \cite{shri}, however, as the Kullback-Leibler divergence $$\sum P[X^{i}_{n}|\vec{{X}}_{0}]\log \frac{ P[X^{i}_{n}|\vec{{X}}_{0}]}{ P[X^{i}_{n}|\vec{{X}}_{0},j]}$$ is only well defined provided $$\{x : P[X^{i}_{n}=x|\vec{{X}}_{0},j]=0\} \subset \{ x: P[X^{i}_{n}=x|\vec{{X}}_{0}]=0\},$$ which is an assumption that does not hold in general. Consider, for example, a directed graph on two vertices $1$ and $2$, with a single directed edge oriented from $1$ to $2$; if we assign the ISD as $\vec{ X}_0=\vec{ e}_1$, then it is trivial that $P[X^{2}_{1}=x|\vec{{X}}_{0}] = \delta_{x}^{1}$ and $P[X^{2}_{1}=x|\vec{{X}}_{0},1]=\delta_{x}^{0}$.\\
Thus to quantify the directed amount of information transferred from vertex $j$ to $i$ we must employ a different measure of statistical distance. There are several possible choices available, among the most promising are the {\it Jensen-Shannon divergence}, which is a linear combination of Kullback-Leibler divergences and the statistical distance introduced by Wootters \cite{Wooters}. Both measures are theoretically rich; Wootter's measure was designed as a distinguishably distance between pure quantum states after a finite number of observations, and applies equally well to distinguishing two probability distributions. The measure also has a geometric interpretation in the context of Hilbert space. The Jensen Shannon Divergence of two distributions, quantifies the total Kullback-Leibler divergence from each distribution to the average of the two, and thus is a measure of distributional similarity. The Jensen Shannon Divergence is also the square of a metric over the space of probability distributions on a measurable set \cite{JSDmet}. These two measures have been shown to agree to second order in a quantum mechanical framework \cite{JSDrev}.\\
We will use the Jensen Shannon Divergence defined by
\begin{eqnarray}
D_{JS}(p,q)=\frac{1}{2}\left(\sum_{x \in \mathcal{X}} \left(p(x) \log \frac{p(x)}{m(x)} + q(x) \log \frac{q(x)}{m(x)}\right)\right)
\end{eqnarray}
where $p,q:\mathcal{X} \rightarrow [0,1]$ are probability distributions (with no restrictions placed on their kernels) and $m=(p+q)/2$. We select this measure as the metric interpretation is of greater use to our theoretical framework.\\
Whence we define the \emph{network transfer entropy} (NTE) from $j$ to $i$ over path length $n$ and given an ISD $\vec{ X}_{0}$ by
\begin{eqnarray}
\tau_{\vec{ X}_0}^{n}(j||i) := D_{JS}(P[X^{i}_{n}|\vec{{X}}_{0}], P[X^{i}_{n}|\vec{{X}}_{0},j])
\end{eqnarray}
This is the central concept of the paper. Note that $\tau_{\vec{ X}_0}^{n}(j||i) \in [0,\log 2]$ is inherently asymmetric, and thus quantifies information transfer through a network in a directed sense, permitting the inference of causality.
\section{Examples}
In order to demonstrate the use of NTE we consider three examples, two synthetic networks and one application to biological signal transduction.
To evaluate NTE in these examples, we estimated the probability distributions $P[X^{i}_{n}|\vec{{X}}_{0}]$ and $P[X^{i}_{n}|\vec{{X}}_{0},j]$ for all $j$, using a simulation. We also devised a method to estimate the statistical error in the probability distributions (see Appendix C).\\
The first and most simple example we consider is a directed path of length 5 with equal edge weights (Figure 1). This induces the stochastic matrix
\begin{eqnarray}
P=\begin{pmatrix}
0 & 1 & 0 & 0 & 0 \\
0 & 0 & 1 & 0 & 0 \\
0 & 0 & 0 & 1 & 0 \\
0 & 0 & 0 & 0 & 1 \\
0 & 0 & 0 & 0 & 1 \end{pmatrix}.
\end{eqnarray}
\begin{figure*}
\centering
\includegraphics{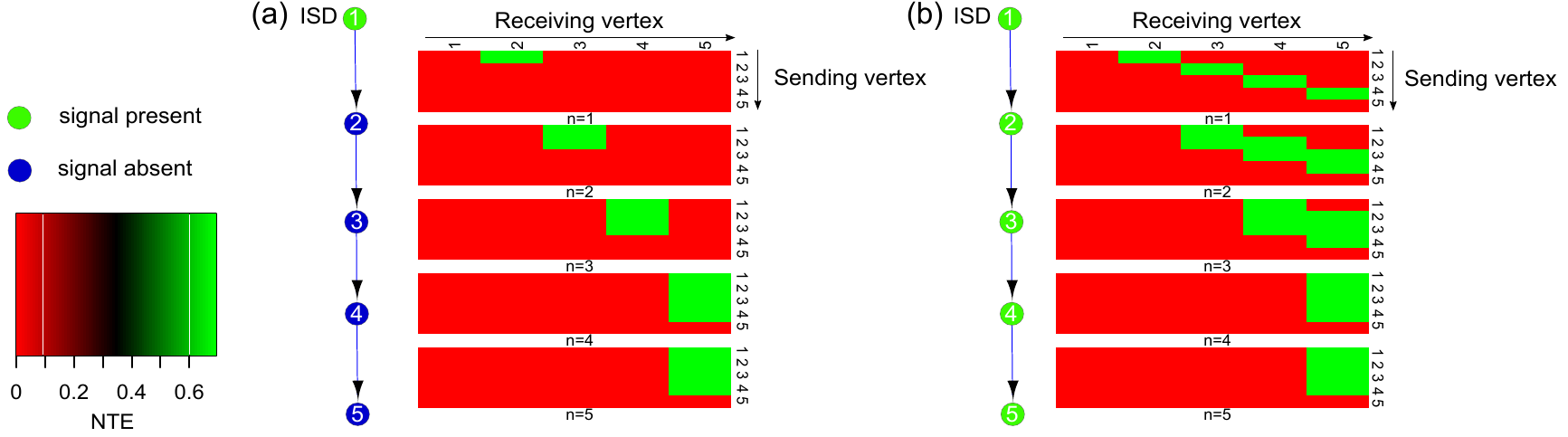}
\caption{(Color Online): Matrices showing NTE between all vertex pairs in a deterministic path over path lengths $n=1-5$ for ISDs $(1,0,0,0,0)^{T}$ (a) and $(1,1,1,1,1)^T$ (b).}
\end{figure*}
The structure of this network provides a completely predictable path for signal transfer and thus is ideal for proving the capability of our measure to detect information transfer. We consider two ISDs on this network, firstly $\vec{ X}_{0}=\vec{ e}_{1}$, where the first vertex in the path is given an initial signal and all other vertices have no signal to transfer. If we number the vertices 1 to 5 from the start of the path to the end, then it is clear that for this ISD, over path length $n=1$, vertex 1 sends information to vertex 2, and no other vertices communicate, for $n=2$ vertex 1 sends information to vertex 3 and vertex 2 sends information to vertex 3 and no other vertices communicate, and similarly we can compute all pairwise information transfer events up to $n=4$ beyond which all signal is absorbed at vertex 5 and cannot be transmitted through the network. This pattern is precisely what is seen if we calculate the NTE between all vertex pairs over different path lengths $n$ (Figure 1).\\
We next consider the ISD $\vec{ X}_{0} = (1,1,1,1,1)^{T}$, on the same network, in order to demonstrate the ability of the NTE measure to discern between situations where networks with identical edge weights have different starting signal states. One would expect that with this ISD, for $n=1$, rather than just vertex 1 forwarding information to vertex 2, we have vertex $j$ forwarding information to vertex $j+1$ for $j=1,\ldots ,4$, and similar extensions for longer path lengths. The NTE measure can detect these differences due to initial signal distribution (Figure 1).\\
The next network we consider is a slight extension to the deterministic path which constitutes a directed feedback from vertex 2 to vertex 4 weighted $w=x/(1-x)$ (Figure 2). This induces the stochastic matrix
\begin{eqnarray}
P=\begin{pmatrix}
0 & 1 & 0 & 0 & 0 \\
0 & 0 & 1 & 0 & 0 \\
0 & 0 & 0 & 1 & 0 \\
0 & x & 0 & 0 & 1-x \\
0 & 0 & 0 & 0 & 1 \end{pmatrix}.
\end{eqnarray}
\begin{figure*}
\centering
\includegraphics{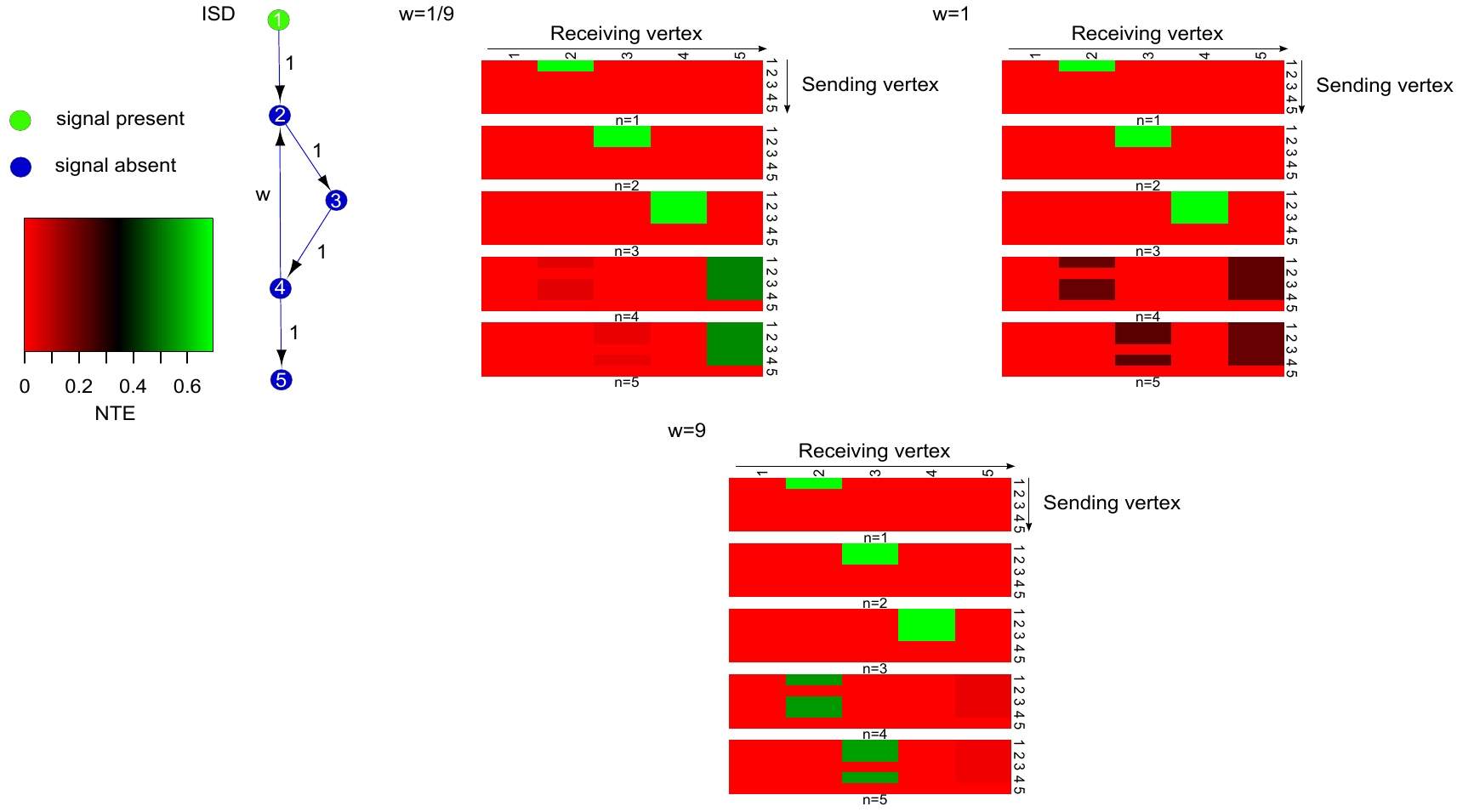}
\caption{(Color Online): Matrices showing NTE between all vertex pairs in the modified, weighted path with feedback from vertex 4 to vertex 2, over path lengths $n=1-5$, for a range of feedback strengths. Note that as the weight on the edge $(4,2)$ rises the NTE from vertices 1-4 to 5 falls.}
\end{figure*}
The network introduces some indeterminism in that if vertex 4 holds signal, it can either forward it to vertex 5 with probability $1-x$ or feedback the signal to vertex 2 with probability $x$. This essentially sets up a feedback loop which dampens the signal received at node 5 over a given path length, by a factor dependant on $x$. We calculated the NTE for all vertex pairs in this altered path for $x=0.1,0.5,0.9$ and found that as $x$ is increased the NTE to vertex 5 from all other vertices falls, as expected (Figure 2). Thus in the context of these very simple synthetic networks, the use of NTE as a tool for detecting information transfer is clear.\\
We next demonstrate NTE in a real world biological network. To do this we consider the human primary naive CD4+ T cell intracellular signalling network analysed by Sachs {\it et al}. \cite{Sachs}, consisting of 11 vertices (Figure 3). In this network vertices are proteins which can be phosphorylated and directed edges connect kinases (capable of phosphorylating proteins) with their targets. The kinases must be in an active state before they can phosphorylate a target; activity can be achieved by either phosphorylation by an upstream kinase or activation by a reagent. Sachs {\it et al}. generated data accompanying this network consisting of quantification (by flow cytometry) of the amount of phosphorylated protein for each vertex in the network following ten independent perturbations. The network perturbations consisted of the administration of reagents which could either activate or inhibit the kinase activity of particular vertices.\\
\begin{figure*}
\centering
\includegraphics{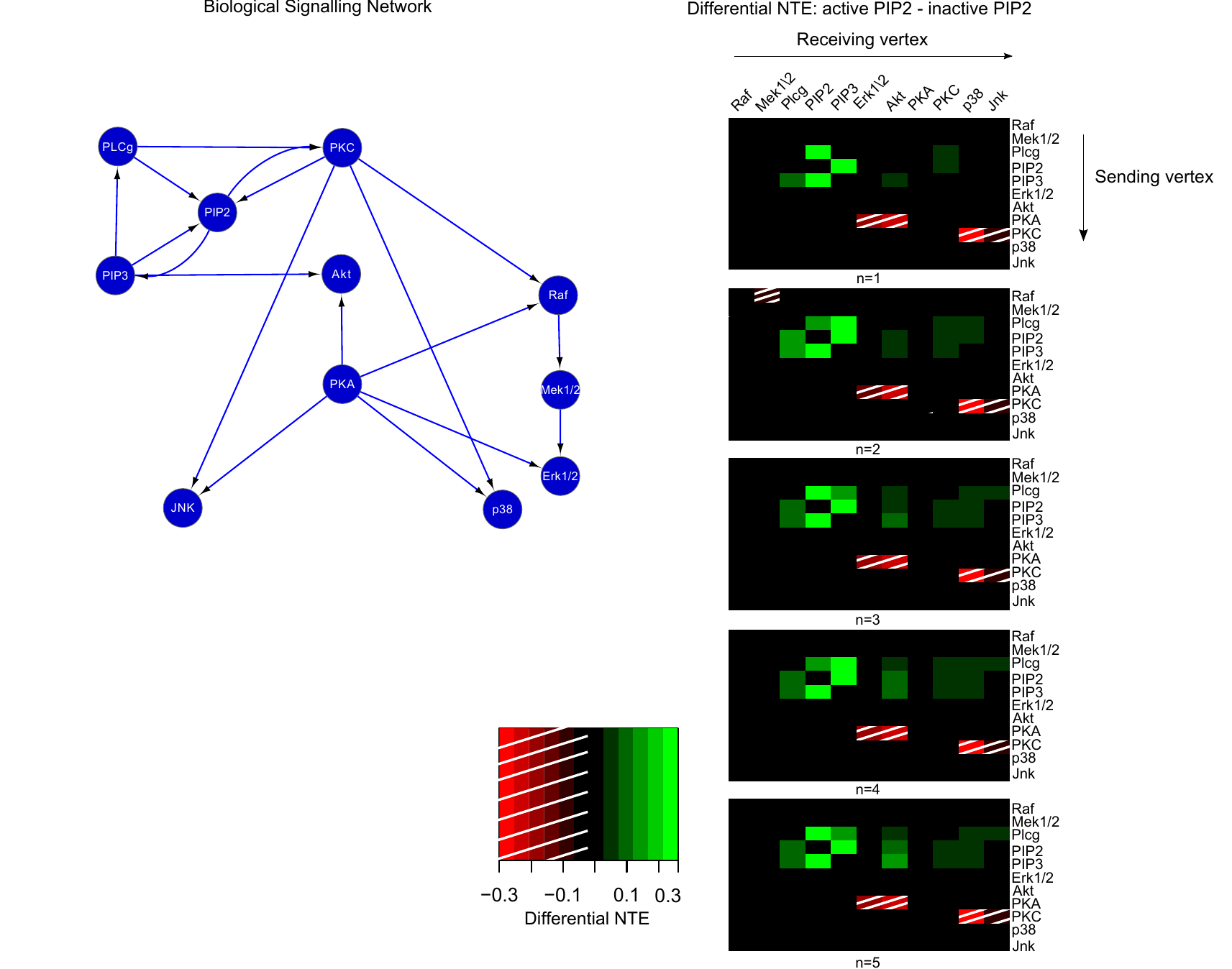}
\caption{(Color Online): Matrix displaying differential NTEs computed for every vertex pair over the displayed network, positive values (light solid, green online) correspond to NTEs higher in the network perturbed with anti-CD3 and anti-CD28, whilst negative values (light dashed, red online) are higher in the network also perturbed by psitectorigenin, a PIP2 inhibitor.}
\end{figure*}
To apply NTE we considered two of these perturbations, firstly treatment with anti-CD3 and anti-CD28 to activate the T-cells and induce flux through the network, secondly treatment with anti-CD3, anti-CD28 and psitectorigenin, a reagent which specifically inactivates PIP2. We computed the NTE between all vertex pairs in the network over paths of length $1-5$ for the two perturbations as described in Appendix D (see Figure 5 in Appendix D, for matrices of NTEs for each perturbation).\\
We found that in the psitectorigenin treated network information transfer from PIP2 to the rest of the network was reduced over all path lengths (Figure 3). Specifically, information transfer from PIP2 to PIP3 was greatly reduced and information transfer from PIP2 to Plcg was reduced over paths of maximal length greater than one (implying PIP3 received less information from Plcg via PIP2 under psitectorigenin treatment). At longer path lengths we also see a reduced information transfer from PIP2 to Akt and p38 in the psitectorigenin treated network. This indicates that specific inhibition of PIP2 can lead to decreased Akt and p38 activation downstream of PIP2 signalling.\\
Interestingly, we also notice that in the PIP2 inhibited network, there is increased information transfer from PKA to Akt and from PKC to p38. This points at a compensatory mechanism, in which inhibition of PIP2 leading to reduced Akt and p38 activation is compensated for by PKC dependant p38 activation and PKA dependant Akt activation.\\
Thus our NTE measure is capable of providing novel insights into signalling mechanisms in biological networks.
\section{A General Framework}
In defining our NTE measure we have additionally constructed a family of mappings from the space of weighted networks to a family of metric spaces, in which elements of the metric spaces correspond to signal dynamics on the networks. The mappings and structure of the metric spaces are parameterised by the path length parameter $n$, the ISD $\vec{ X_{0}}$ and the topology ({\it i.e.,} the zero pattern, but not the edge weights) of the network and their construction is explained in detail in Appendix B.\\
This formalism allows a more theoretical treatment of dynamics on networks from the perspective of metric spaces, and permits a coupling between weighted network structure and dynamics. In certain fields, understanding the reaction of network dynamics to perturbations of edge weights is of great importance. This is particularly true of network drug design \cite{netdrugrev}, in which one is interested in sequentially deforming the quantitative strengths of interactions in a pathological signalling network (via drugs) into those of a healthy network, with the aim of establishing a healthy gene expression dynamic and mitigating the pathology. If this notion of treatment is logical within our framework, then one would postulate that convergence in weight distribution of a sequence of networks to a limit distribution (in a matrix metric space) would imply convergence of the corresponding sequence of dynamics to the dynamics of the limit network (in the network dynamic metric space). We state and prove a theorem in Appendix E which establishes this postulate as true. This result demonstrates an intuitive coupling between network structure and dynamics within our framework.\\
Further theoretical questions may consider which ISDs are maintained under different networks, these represent persistent (attractor) states of the network information distribution. To identify such states we note that every graph (if we permit self edges at every vertex to represent a non-zero probability of signal maintenance) admits a disjoint vertex cycle decomposition \cite{vertexcycle}. Thus there is always a way of sending signal around the network, without combining signal from two vertices at any one vertex. This implies that for every weighted network $W$, with self edges, there must exist a permutation matrix $\phi$, which admits at least one vector $vec{x}$ satisfying $\phi \vec{x} = \vec{x}$ ({\it e.g.,} the vector $(1,\ldots ,1)^{T}$), such that $P[X_{1}=\phi \vec{x}|\vec{{X}}_{0}={ \vec{x}}]>0$. The state $\vec{x}$ thus has a non-zero probability of being a persistent information state of the network.\\
Questions concerning the evolution of self-assembling networks can also be considered in our framework via an application of dynamic programming. An introduction to this approach is detailed in Appendix F.
\section{Conclusion}
We have derived a general information theoretic measure, network transfer entropy, for quantifying the amount of information transferred between any two vertices of a weighted network over paths of varying length. We have demonstrated our measure on simple synthetic weighted networks and applied it to biological signal transduction, revealing insights into the robustness of kinase signalling. We have also constructed a general metric space framework for dynamics on weighted networks and proved a convergence principle relating weighted network structure to dynamics. We outlined how problems in network evolution and network dynamic stability can be approached within our framework, formalisation of these approaches is a topic of future work.

\appendix
\section{Introduction}
In this appendix we provide certain mathematical derivations and technical methodologies to accompany the main text. In Appendix B we derive a closed form expression for the probability distribution $P[X_{n}^{i}|\vec{ X_{0}}]$, describing the signal at a vertex $i$ in a weighted network, given an initial signal distribution (ISD), $\vec{ X}_0$, has traversed a path of length $n$. Following the derivation of this expression we will explain how the network transfer entropy (NTE) framework leads to the construction of a family of mappings from the space of weighted networks to a family of metric spaces describing signal dynamics. In Appendix C we explain how the NTE may be estimated from a simulation of the Markovian dynamic model introduced in the main text and how error may be compensated for in this estimation. In Appendix D we explain in detail the application of NTE to biological signal transduction, specifically the assignment of an ISD and edge weights for each perturbation. In Appendix E we state and prove aa convergence theorem in the metric space framework. Finally, in Appendix F, we outline an approach to network evolution from the perspective of dynamic programming.
\section{$P[X_{n}^{i}|\vec{ X_{0}}]$ and metric space}
To compute the probability distribution $P[X_{n}^{i}|\vec{ X_{0}}]$ for a given weighted network $W=(w_{ij})_{i,j=1}^{N}$, with corresponding stochastic matrix $P=(p_{ij})_{i,j=1}^{N}$ (see main text) we first note that
\begin{eqnarray}\label{2}
P[X_{n}^{i}=y|\vec{ X_{0}}] = \sum_{\vec{ x}} P[\vec{ X}_{n}=\vec{ x}|\vec{ X_{0}}]\delta_{x_{i}}^{y},
\end{eqnarray}
Where $\delta_{i}^{j}$ denotes the Kronecker delta of $i$ and $j$.\\
In addition by the Markovian nature of our dynamic model
\begin{widetext}\begin{eqnarray}
\label{1}
P[\vec{ X}_{n}=\vec{{x}}|\vec{ X_{0}}]= \sum_{\vec{ X}_{1},\ldots ,\vec{ X}_{n-1}}P[\vec{ X}_{n}=\vec{{x}}|\vec{ X}_{n-1}]\ldots P[\vec{ X}_1|\vec{ X}_{0}],
\end{eqnarray}\end{widetext}
Reducing our problem to the calculation of the transition probabilities $P[\vec{ X}_{k+1}|\vec{ X_{k}}]$, between states and the states themselves which must be summed over. These are not, however, immediate. For the calculation consider the following: given we know the full signal distribution at time-point $k \in \mathbb{N}$ {\it{i.e.}} $\vec{{X}}_{k} = \vec{{x}}_{k}$, then all possible states of signal distribution at time-point $k+1$ have the form
\begin{eqnarray}
\vec{{X}}_{k+1} = A^{T}\vec{{x}}_{k}.
\end{eqnarray}
Here $A = (A_{ij})_{i,j=1}^N$, $A_{ij} \in \{0,1\}$ is a binary matrix with a single non-zero entry in every row; the column index $j$ of the non-zero entry in row $i$ corresponds to the unique vertex $j$ that $i$ has sent its signal to during the time-step $k \rightarrow k+1$. We note that in addition $A_{ij} = 0$ if $j \not \in \mathcal{N}_i$ and that $A$ is independent of $\vec{ x}_{k}$.\\
Thus every realisation of a single signal transfer event in a given weighted network can be represented as a matrix operation $A$, independently of ISD. We denote the set of such matrices by $\mathcal{A}$, and emphasise that it depends only upon the topology of the weighted network.\\
It is clear that for $N<\infty$ the set $\mathcal{A}$ must be countable, and its cardinality must be $|\mathcal{A}| = \prod_{i=1}^{N} k_i$, where $k_i = |\mathcal{N}_{i}|$. Moreover, it is clear that we can construct every element in $\mathcal{A}$ given $\cup_{i=1}^{N} \mathcal{N}_{i}$.\\
Following this, it is clear that given any ISD $\vec{ X}_0$ the signal distribution at time point $k>0$ must have the (possibly non-unique) form
\begin{eqnarray}
\vec{{X}}_{k} = A_{k}^{T}\ldots A_{1}^{T}\vec{{X}}_{0},
\end{eqnarray}
where $A_{i} \in \mathcal{A}$ for $i=1,\ldots ,k$. Whence Eq. \eqref{1} can be expressed as
\begin{widetext}\begin{eqnarray}\label{3}
P[\vec{ X}_{n}=\vec{{x}}|\vec{ X_{0}}]= \sum_{A_{1},\ldots ,A_{n-1} \in \mathcal{A}}P[\vec{ X}_{n}= \vec{ x}|\vec{ X}_{n-1}= A_{n-1}^{T}\ldots A_{1}^{T}\vec{{X}}_{0}]\ldots P[\vec{ X}_1=A^{T}_{1}\vec{ X}_{0}|\vec{ X}_{0}].
\end{eqnarray}\end{widetext}
Thus to compute $P[X_{n}^{i}|\vec{ X_{0}}]$, it suffices to compute $$P[\vec{ X}_{k+1}= A_{k+1}^{T}\ldots A_{1}^{T}\vec{{X}}_{0}|\vec{ X}_{k}= A_{k}^{T}\ldots A_{1}^{T}\vec{{X}}_{0}],$$ which is simply the probability of the signal dynamic $A_{k+1}$ being selected from $\mathcal{A}$. By model construction this can be expressed as $$\prod_{i=1}^{N} \sum_{j=1}^{N} p_{ij}A_{ij}.$$ Whence combining this with \eqref{2} and \eqref{3} we derive the closed form expression
\begin{widetext}\begin{eqnarray}
P[X_{n}^{i}=y|\vec{ X_{0}}] = \sum_{A_{1},\ldots ,A_{n} \in \mathcal{A}} \delta_{ ({A_{1}^{T}\ldots A_{n}^{T}\vec{ X}_0})_{i}}^{y} \prod_{r=1}^{n} \prod_{k=1}^{N} \sum_{j=1}^{N} p_{kj}{(A_{r}})_{kj}.
\end{eqnarray}\end{widetext}
\subsection{Metric Space}
We demonstrated above how, for a specific network $W$, and ISD $\vec{ X}_{0}$ we can calculate a set of matrices describing possible signal dynamics over a single time-step of our model, as well as a probability distribution describing the signal on the entire network after $n$ time steps. We will denote these constructs for the weighted network $W$ by $\mathcal{A}_{W}$ and $P_{W}[\vec{ X}_{n}|\vec{{X}}_{0}]$, respectively and stress that the former only depends on the topology of $W$ and not the edge weights, we will denote the topology of $W$ by $t(W)$ (topology in this context refers to the zero pattern of the network and is independent of the edge weights). The probability distribution $P_{W}[\vec{ X}_{n}|\vec{{X}}_{0}]$ is a measure over the finite set $$\{A_{1}^{T}\ldots A_{n}^{T}\vec{{X}}_0: (A_{i})_{i=1}^{n} \subset \mathcal{A}_W\},$$ which we will denote $\Omega_{\vec{ X}_{0}}^{n}(t(W))$. If we denote the space of probability measures over $\Omega_{\vec{ X}_{0}}^{n}(t(W))$ by $M^{+}_{1}(\Omega_{\vec{ X}_{0}}^{n}(t(W)))$, then it is clear that for any two weighted networks $W_{1}$ and $W_{2}$ with the same topology, $T$, the probability distributions $P_{W_{1}}[\vec{ X}_{n}|\vec{{X}}_{0}]$ and $P_{W_{2}}[\vec{ X}_{n}|\vec{{X}}_{0}]$ are elements of $M^{+}_{1}(\Omega_{\vec{ X}_{0}}^{n}(T))$. \\
It has been shown that for any measurable space $\Omega$, the square root of the Jensen Shannon Divergence induces a metric on the space $M^{+}_{1}(\Omega)$ \cite{JSDmet}, thus the quantity $$\sqrt{D_{JS}(P_{W_{1}}[\vec{ X}_{n}|\vec{{X}}_{0}],P_{W_{2}}[\vec{ X}_{n}|\vec{{X}}_{0}])}$$ computes a metric distance between the probability distributions describing the dynamics on $W_{1}$ and $W_2$.\\
Thus our NTE framework results in a mapping from the space of weighted networks to a family of metric spaces in which elements of the  metric space represent possible signal dynamics.
\section{Estimating NTE}
Network transfer entropy is formulated as the Jensen Shannon Divergence between two probability distributions. As explored above we can derive closed form expressions for these probability distributions, however, their evaluation can be computationally expensive, if there are multiple vertices of large degree. This is because a main step in the evaluation of the expressions is constructing the set $\mathcal{A}$ of possible signal dynamics over a single time step, which for a network on $N$ vertices is of dimension $\prod_{i=1}^{N} k_{i}$. Moreover, the time complexity of evaluation scales exponentially with the path length parameter $n$.\\
For most networks, however, estimation of the probability distributions involved in the NTE expression can be done efficiently. As the model underlying these distributions is a discrete time Markov chain, with a discrete state space, we can employ Monte Carlo simulation for any ISD to provide realisations of the signal distribution on the entire network, for any path length parameter $n$. From these realisations the probability of a specified signal level at vertex $i$, given an ISD and path length parameter $n$, can be estimated as the proportion of simulations in which the specified level is achieved.\\
Two major considerations need to be addressed to ensure accurate estimation from this procedure. Firstly, it is clear that the more simulations of the model performed, the more accurate the estimate of the probability distribution, moreover the estimate computed from $K$ simulations will converge to the true distribution as $K \rightarrow \infty$. Thus it is essential to select $K$ sufficiently large to ensure that the estimated distribution is sufficiently near the true distribution with high probability. Secondly, given a specified $K$ it is important to establish how the error in estimating the probability distributions translates to error in estimating the NTE.\\
To address the first issue, we consider only the full network {\it i.e.,} without any vertices set to absorbing state, as the stochastic matrix for the full network will have the fewest deterministic vertices, and thus will be the hardest to estimate probability distributions for. For each probability to be estimated we construct a trace plot describing the change of the estimate with the number of simulations $K$. This plot allows us to assess convergence of the estimate as $K$ is increased. We select the number of simulations $K$ for each network as the maximal $K$ such that, the shape of the trace plots indicates convergence and the estimates (for every vertex signal probability) at $K$ and $K-100$ differ by no more than $0.01$.\\
To address the second issue of error in the NTE after selecting $K$, we computed multiple ($R$) estimations of the signal probability distributions for the full network from $K$ simulations. We then computed, for every ${R}\choose{2}$ estimate pair, the Jensen Shannon Divergence between the two estimates of the signal distributions at each vertex. This Jensen Shannon Divergence, computed for vertex $i$, can be interpreted as the NTE from a vertex $j$ to $i$, when $j$ sends no information to $i$, if the estimation is perfect, this quantity should be zero. As the estimation is imperfect we obtain ${R}\choose{2}$ estimates of the error in the NTE, deriving from error in the estimation of the probability distributions from simulation, for each receiving vertex. From these estimates we can estimate the first two moments of the error distribution, a NTE for a given receiving vertex is considered not attributable to error, provided it lies at least two standard deviations above the maximal error observed for the vertex.
\section{Computing the ISD and edge weights for the biological network}
To compute the NTEs over the two perturbations of the biological signalling network considered in the main text we must first define the ISD and the edge weights from the data. As the kinases in the network must be phosphorylated to phosphorylate their direct targets, two connected proteins with highly positively correlated phosphorylation levels across single cell observations under a given perturbation, are likely interacting. Thus a suitable edge weight which captures the strength of a phosphorylation interaction represented by an edge $(i,j)$ under a given perturbation $k$ is $1+C^{k}_{ij}$, where $C_{ij}^{k}$ is the Pearson correlation of phosphorylated protein levels across single cell measurements under perturbation $k$.\\
Defining the ISD is less trivial and requires consideration of the question we wish to answer and some technicalities. To determine the differences in information transfer between the two perturbations, it makes sense to consider an ISD which quantifies the difference in phosphorylated protein levels between the two treatments.  A technical issue to consider is how different signal distributions on the same network lead to different NTE values between vertices. It is clear that weighting a vertex with a non-zero signal leads to a higher NTE value between that vertex and its downstream interaction partners (depending on the value of $n$) than weighting the vertex with zero signal. Thus vertices with an information deficit in one perturbation versus another should be weighted with zero signal in that perturbation and a non-zero signal in the other. \\
A more subtle issue concerns the number of unique signal values attainable at each vertex for a given ISD and how this relates to the NTE. It is somewhat intuitive that if the inputs to a given vertex each have a unique initial signal value, then the range of values attainable by the receiving vertex will be more diverse than if all the inputs had the same initial signal value. Thus one may hypothesise that the NTE from one vertex to another vertex, with multiple inputs, will be larger if the input vertices have unique initial signal values than if they have identical signal values.\\
To explain this concept, consider the network shown in Figure 4, consisting of one central vertex with 3 possible inputs, which are each as likely to forward signal to central vertex as they are to forward signal to a separate independent neighbour.
\begin{figure*}
\centering
\includegraphics{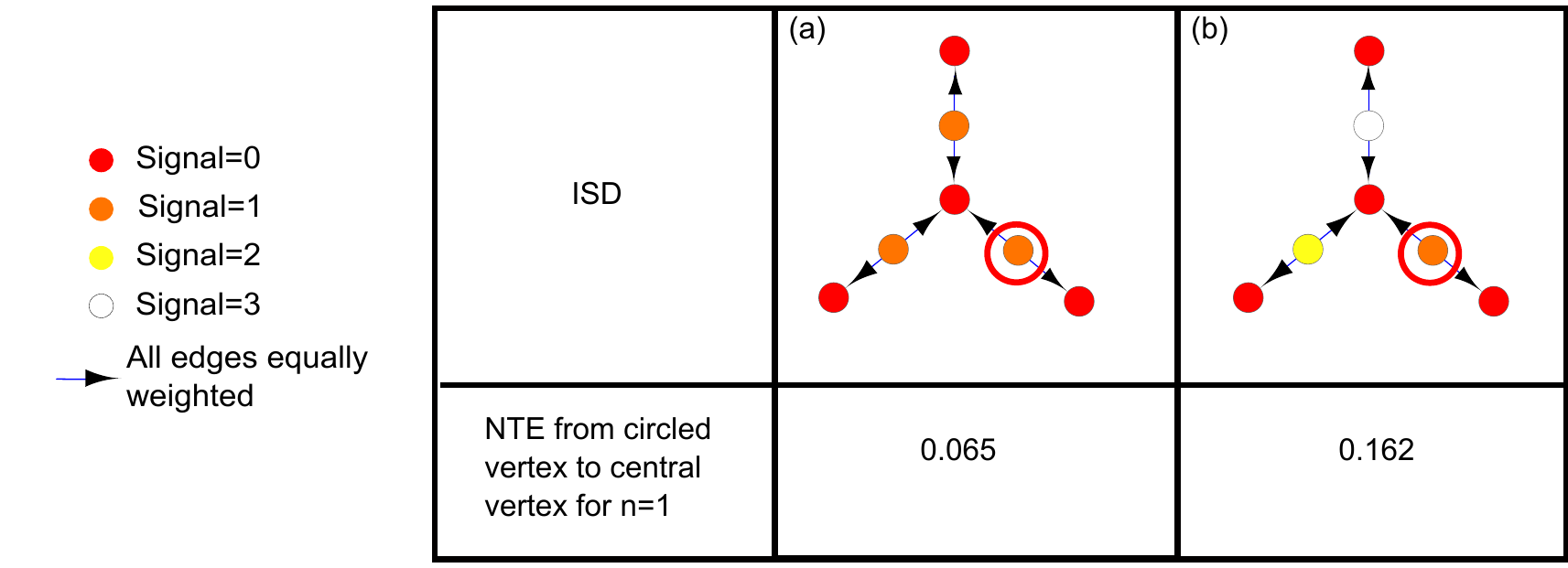}
\caption{(Color Online): Comparing the NTE from source vertices to a target vertex when the ISD at source vertices are non-unique (a) and unique (b), note that a unique ISD at input vertices results in a higher NTE from source to target.}
\end{figure*}
We consider the effect of the ISD on the NTE from the circled vertex to the central vertex for path length parameter $n=1$ via two cases. In the first scenario the initial signal at every input vertex is identical (in this example this signal value is one), while the signal at every other vertex starts at zero. In the second case every input vertex is given a unique value (in our example these values are simply 1, 2 and 3), while the other vertices are again given initial signals of 0. Consider the probability distribution of signal at the central vertex after a single signal transfer event. For the first ISD, this distribution can take 4 unique values (namely 0, 1, 2 and 3), however for the second ISD, with unique signal values at the inputs, the distribution can take 7 values (integers 0 through 6). Now consider the signal distribution at the central vertex given that we prevent the circled vertex from sending information. For the first ISD, this distribution now admits only 3 possible values (0, 1 and 2), while for the second ISD only 4 possible values are now attainable (0, 2, 3 and 5). Thus the size of the ``coding alphabet" at the central vertex after removal of input from the circled vertex has shrunk from 4 to 3 under the first ISD but from 7 to 4 under the second ISD (a much greater fall). Consequentially, we notice that the NTE from the circled vertex to the central vertex is lower for the first ISD than the second ISD.\\
If we follow this concept that inputs with unique initial signal send more information to their outputs, it is logical that vertices with significantly higher signal in one perturbation versus another should be given a unique initial signal value in the first perturbation, reflecting their capacity to send more information about the network.\\
All that remains now is to consider how to assign initial signal to vertices which do not display a great difference in signal distribution across the two perturbations. One solution to this problem is to assign all these vertices an identical initial signal, in this way they can transfer more information than vertices with a signal deficit in one perturbation versus another but less information than vertices with a signal surplus.\\
Guided by these concepts we constructed the ISD for each perturbation as follows. We utilised the limma package in R \cite{limma} to compute t-values testing, for each vertex in the network, the hypothesis that the phosphorylated protein level of the vertex was significantly different in the two treatments. If for a given vertex the phosphorylated protein levels were significantly lower ($p<0.05$) in one perturbation versus another it was assigned an initial value of zero in that perturbation and a unique initial value (here chosen as the absolute t-statistic of the test) in the other perturbation. All vertices which did not display significant changes between the two perturbations were assigned the same non-unique initial value of 1 in both perturbations.\\
The ISDs and edge weights for the two perturbations alongside NTE matrices computed for each perturbation are provided in Figure 5.
\section{Proof of the convergence principle}
In this appendix we prove the following theorem
\begin{theorem}[Convergence Principle]
Let $(W^n)_{n \in \mathbb{N}}$ be a sequence of weighted networks on $N$ vertices of fixed topology, $T$, and let $(P^{m})_{m \in \mathbb{N}} \subset [0,1]^{N \times N}$ be the corresponding row normalised stochastic matrices for the sequence. Let $P \subset [0,1]^{N \times N}$ be a stochastic matrix of topology $T$. If $P^m \rightarrow P$ in $L^{p}(\mathbb{M}^{N \times N})$, $p\geq 1$, as $m \rightarrow \infty$, then for fixed ISD $\vec{ X}_{0}$ and path length parameter $n$, the signal distributions $$P_{P^{m}}[\vec{ X}_{n}|\vec{{X}}_{0}] \rightarrow P_{P}[\vec{ X}_{n}|\vec{{X}}_{0}]$$ as $m \rightarrow \infty$ in the metric space $$\big(M^{+}_{1}(\Omega_{\vec{ X}_{0}}^{n}(T)),\sqrt{D_{JS}(\cdot,\cdot)}\big).$$
\end{theorem}
\subsection{Proof}
Firstly we define the $L^{p}$ norm of a matrix $A \in \mathbb{M}^{N\times N}$
\begin{eqnarray}
||A||_{p} :=   \begin{cases}
\left(\sum_{i,j=1}^{N} |a_{ij}|^{p}\right)^{1/p}
& \text{if $p<\infty$} \\
\max_{i,j} |A_{ij}|    & \text{if $p=\infty$}
\end{cases}.
\end{eqnarray}
A well-known and easy to derive bound on $L^{p}$ spaces, which holds for any $A \in \mathbb{M}^{N \times N}$ is $||A||_{\infty} \leq ||A||_{p}.$\\
Fix $\epsilon>0$. As $P^m \rightarrow P$ in $L^{p}(\mathbb{M}^{N \times N})$, we have that: there exists $M \in \mathbb{N}$ such that for all $m\geq M$, $$||P^{m}-P||_{p}\ < \epsilon.$$\\
Let us define the matrix $\Delta^{P} \in (-1,1)^{N \times N}$ via $$\Delta^{P} := P^M -P,$$ it is clear that $||\Delta^{P}||_{\infty}<\epsilon$.\\
We now consider for a fixed ISD $\vec{ X}_{0}$ and path length parameter $n$ the distributions $P_{P}[\vec{ X}_{n}=\vec{x}|\vec{{X}}_{0}]$ and $P_{P^{M}}[\vec{ X}_{n}=\vec{x}|\vec{{X}}_{0}]$, which we will hereafter refer to as $\mathbb{P}_{P}(\vec{x})$ and $\mathbb{P}_{P^M}(\vec{x})$. It was shown above that
\begin{widetext}\begin{eqnarray}\label{b}
\mathbb{P}_{P}(\vec{x}) = \sum_{A_{1},\ldots ,A_{n} \in \mathcal{A}} \delta_{ ({A_{1}^{T}\ldots A_{n}^{T}\vec{ X}_0})}^{\vec{x}} \prod_{r=1}^{n} \prod_{i=1}^{N} \sum_{j=1}^{N} P_{ij}{(A_{r}})_{ij}.
\end{eqnarray}\end{widetext}
The set of possible signal dynamics $\mathcal{A}$ for a given weighted network was also explicitly constructed above and was shown to depend only on the network topology and not on the edge weights of the network. Consequentially as the sequence $(P^{n})_{n \in\mathbb{N}}$ and the network $P$ have the same topology $T$, the set of possible signal dynamics $\mathcal{A}$ is the same for every element of the sequence and the network $P$.\\
Consider expanding the product
\begin{widetext}
\begin{eqnarray}
\prod_{r=1}^{n} \prod_{i=1}^{N} \sum_{j=1}^{N} P_{ij}{(A_{r}})_{ij} &=& \prod_{r=1}^{n} \prod_{i=1}^{N} \sum_{j=1}^{N} (P^{M}_{ij}-\Delta^{P}_{ij}){(A_{r}})_{ij}\\
&=& \Bigg[ \big(\sum_{j=1}^{N} (P^{M}_{1j}(A_{1})_{1j}
-\sum_{j=1}^{N} \Delta^{P}_{1j})(A_{1})_{1j}\big)
\ldots \big(\sum_{j=1}^{N} (P^{M}_{Nj}(A_{1})_{Nj}
-\sum_{j=1}^{N} \Delta^{P}_{Nj})(A_{1})_{Nj}\big)\Bigg]\notag\\
&\ldots&  \Bigg[\big(\sum_{j=1}^{N} (P^{M}_{1j}(A_{n})_{1j}
-\sum_{j=1}^{N} \Delta^{P}_{1j})(A_{n})_{1j}\big)
\ldots  \big(\sum_{j=1}^{N} (P^{M}_{Nj}(A_{n})_{Nj}
-\sum_{j=1}^{N} \Delta^{P}_{Nj})(A_{n})_{Nj}\big)\Bigg].
\end{eqnarray}\end{widetext}
Grouping together terms we can express the product as
\begin{widetext}\begin{eqnarray}\label{a}
\prod_{r=1}^{n} \prod_{i=1}^{N} \sum_{j=1}^{N} P_{ij}(A_{r})_{ij} &=& \prod_{r=1}^{n} \prod_{i=1}^{N} \sum_{j=1}^{N} P^{M}_{ij}{(A_{r}})_{ij} \notag\\
&-& \Bigg[ \sum_{i=1}^{N} \sum_{r=1}^{n} \big(\sum_{j=1}^{N} \Delta^{P}_{ij}(A_{r})_{ij}\big)
\prod_{l\not =r}^{n} \prod_{s \not =i}^{N} \big(\sum_{j=1}^{N} P^M_{sj}{(A_{l}})_{sj}\big)\Bigg] + o(\epsilon).
\end{eqnarray}\end{widetext}
We will denote the second term in the above expression by
\begin{widetext}\begin{eqnarray}
\mathcal{H}_{(A_{i})_{i=1}^{n}}(\Delta^{P}):=\Bigg[ \sum_{i=1}^{N} \sum_{r=1}^{n} \big(\sum_{j=1}^{N} \Delta^{P}_{ij}(A_{r})_{ij}\big) \prod_{l\not =r}^{n} \prod_{s \not =i}^{N}\big( \sum_{j=1}^{N} P^M_{sj}{(A_{l}})_{sj}\big)\Bigg]
\end{eqnarray}\end{widetext}
Substitution of \eqref{a} into \eqref{b} yields
\begin{widetext}\begin{eqnarray}\label{c}
\mathbb{P}_{P}(x) = \sum_{A_{1},\ldots ,A_{n} \in \mathcal{A}} \delta_{ ({A_{1}^{T}\ldots A_{n}^{T}\vec{ X}_0})}^{\vec{x}}\Bigg[\prod_{r=1}^{n} \prod_{i=1}^{N}
\sum_{j=1}^{N} P^{M}_{ij}{(A_{r}})_{ij} - \mathcal{H}_{(A_{i})_{i=1}^{n}}(\Delta^{P}) + o(\epsilon)\Bigg].
\end{eqnarray}\end{widetext}
Clearly from \eqref{b} the first term can be expressed
\begin{widetext}\begin{eqnarray}
\sum_{A_{1},\ldots ,A_{n} \in \mathcal{A}} \delta_{ ({A_{1}^{T}\ldots A_{n}^{T}\vec{ X}_0})}^{\vec{x}}\prod_{r=1}^{n} \prod_{i=1}^{N} \sum_{j=1}^{N} P^{M}_{ij}{(A_{r}})_{ij} =\mathbb{P}_{P^{M}}(x).
\end{eqnarray}\end{widetext}
For the second term, notice that\newpage
\begin{widetext}\begin{eqnarray}\label{e}
\left|\sum_{A_{1},\ldots ,A_{n} \in \mathcal{A}} \delta_{ ({A_{1}^{T}\ldots A_{n}^{T}\vec{X}_0})}^{\vec{x}}\mathcal{H}_{(A_{i})_{i=1}^{n}}(\Delta^{P})\right| &\leq& \sum_{A_{1},\ldots ,A_{n} \in \mathcal{A}} \delta_{({A_{1}^{T}\ldots A_{n}^{T}\vec{ X}_0})}^{\vec{x}}\sum_{i=1}^{N} \sum_{r=1}^{n} \big(\sum_{j=1}^{N}||\Delta^{P}||_{\infty}(A_{r})_{ij}\big)\notag\\
&&\prod_{l\not =r}^{n} \prod_{s \not =i}^{N}\big( \sum_{j=1}^{N} P^M_{sj}{(A_{l}})_{sj}\big)\notag\\
&<& \epsilon\sum_{A_{1},\ldots ,A_{n} \in \mathcal{A}} \delta_{ ({A_{1}^{T}\ldots A_{n}^{T}\vec{ X}_0})}^{x}\sum_{i=1}^{N} \sum_{r=1}^{n}\prod_{l\not =r}^{n} \prod_{s \not =i}^{N}\big( \sum_{j=1}^{N} P^M_{sj}{(A_{l}})_{sj}\big)\notag\\
&\leq& \left(\prod_{i=1}^{N}k_{i}\right)^{n}n N\epsilon
\end{eqnarray}\end{widetext}
where the second inequality follows from $||\Delta^{P}||_{\infty}<\epsilon$ and $\sum_{j=1}^{N}(A_{r})_{ij}=1$ by construction of the set $\mathcal{A}$. The final inequality follows from the facts that $\sum_{j=1}^{N} P^M_{sj}{(A_{l}})_{sj} \leq 1$ and $|\mathcal{A}|=\prod_{i=1}^{N}k_{i}$.
Given these bounds and Eq. \eqref{c} it follows that
\begin{eqnarray}\label{d}
\mathbb{P}_{P}(x) < \mathbb{P}_{P^{M}}(x)+ \left(\prod_{i=1}^{N}k_{i}\right)^{n} nN\epsilon + o(\epsilon).
\end{eqnarray}
An identical argument can be used exchanging $P^{M}$ and $P$, in which case the sign of $\mathcal{H}_{(A_{i})_{i=1}^{n}}(\Delta^{P})$ in \eqref{c} changes to positive, however the bound established in \eqref{e} bounds the modulus of $\mathcal{H}_{(A_{i})_{i=1}^{n}}(\Delta^{P})$ and thus will always be greatest than the largest negative or largest positive value of $\mathcal{H}_{(A_{i})_{i=1}^{n}}(\Delta^{P})$. Thus we obtain the symmetric bound
\begin{eqnarray}\label{f}
\mathbb{P}_{P^{M}}(x) < \mathbb{P}_{P}(x) + \left(\prod_{i=1}^{N}k_{i}\right)^{n} nN\epsilon + o(\epsilon).
\end{eqnarray}
Let us define $$m(x):=\frac{1}{2}\left(\mathbb{P}_{P}(x)+\mathbb{P}_{P^{M}}(x)\right),$$ it follows from \eqref{d} and \eqref{f} that
\begin{eqnarray}
\frac{\mathbb{P}_{P}(x)}{m(x)} < \frac{2\mathbb{P}_{P}(x)}{2\mathbb{P}_{P}(x)-\left(\prod_{i=1}^{N}k_{i}\right)^{n}nN\epsilon +o(\epsilon)}
\end{eqnarray}
and
\begin{eqnarray}
\frac{\mathbb{P}_{P^{M}}(x)}{m(x)} < \frac{2\mathbb{P}_{P^{M}}(x)}{2\mathbb{P}_{P^{M}}(x)-\left(\prod_{i=1}^{N}k_{i}\right)^{n}nN\epsilon +o(\epsilon)}.
\end{eqnarray}
Thus it follows that
\begin{widetext}\begin{eqnarray}
D_{JS}(\mathbb{P}_{P},\mathbb{P}_{P^{M}}) &=& \frac{1}{2}\Bigg(\sum_x \mathbb{P}_{P}(x) \log \frac{\mathbb{P}_{P}(x)}{m(x)} + \mathbb{P}_{P^{M}}(x) \log \frac{\mathbb{P}_{P^{M}}(x)}{m(x)}\Bigg)\notag\\
&<& \frac{1}{2}\Bigg(\sum_{x} \log \left(\frac{2\mathbb{P}_{P}(x)}{2\mathbb{P}_{P}(x)-\left(\prod_{i=1}^{N}k_{i}\right)^{n}nN\epsilon +o(\epsilon)}\right)\notag\\&+&\log\left(\frac{2\mathbb{P}_{P^{M}}(x)}{2\mathbb{P}_{P^{M}}(x)-\left(\prod_{i=1}^{N}k_{i}\right)^{n} nN\epsilon +o(\epsilon)}\right)\Bigg).
\end{eqnarray}\end{widetext}
By algebra of limits, it is clear that as $\epsilon \rightarrow 0$ $$\frac{2\mathbb{P}_{P}(x)}{2\mathbb{P}_{P}(x)-\left(\prod_{i=1}^{N}k_{i}\right)^{n}nN\epsilon +o(\epsilon)}\rightarrow 1$$ and $$\frac{2\mathbb{P}_{P^{M}}(x)}{2\mathbb{P}_{P^M}(x)-\left(\prod_{i=1}^{N}k_{i}\right)^{n}nN\epsilon +o(\epsilon)}\rightarrow 1.$$ Whence it follows that
$$D_{JS}(\mathbb{P}_{P},\mathbb{P}_{P^{m}})\rightarrow 0$$ as $m \rightarrow \infty$ and the theorem is true.\\
We note that the theorem also holds if the topologies of the sequence of weighted networks and limit network are different from one another, the proof of this statement follows precisely as above, the only difference being the set $\mathcal{A}$ utilised is that induced by the complete graph topology.
\section{Network evolution and dynamic programming}
As mentioned in the main text the evolution of self-assembling networks can also be considered in our framework via an application of dynamic programming. To see this we consider the space $\mathcal{A}$ (explicitly constructed above) containing matrix representations of all possible single path length, signal forwarding choices, induced by the complete graph on $K$ vertices. We note that for every weighted network, $W$, on $N$ vertices, where $N<K$, the corresponding stochastic matrix $P$ can be expressed as a convex combination of elements in $\mathcal{A}$, $P = \sum_{j=1}^{K^N} \rho_j A_{j}$ where $\{A_{j}\}_{j=1}^{K^N} = \mathcal{A}$, $\sum_{j}\rho_j = 1$, $\rho_{j} \leq 1$ for all $j$. If one interprets the space $\mathcal{A}$ as a state space of possible choices of signal dynamics through the network and considers $\vec{ \rho} =\{\rho_j\}_{j=1}^{K^N}$ as a policy, giving a probability distribution of selecting a given global signal dynamic from the state space, that has been obtained by some optimality criterion, then one has a dynamic programming framework for network dynamic evolution. We note that one can calculate the policy explicitly, as $\rho_{i} = P[\vec{ X}_{1}=A_{i}\vec{ X}_{0}|\vec{ X}_{0}]$. Thus we have information to guide to construction of an optimality criterion describing network evolution. Forms of such a criterion can be posited and parameterised for different systems and suitable parameter regimes can be reverse engineered from the policy solution $\vec{\rho}$.
\begin{figure*}
\centering
\includegraphics{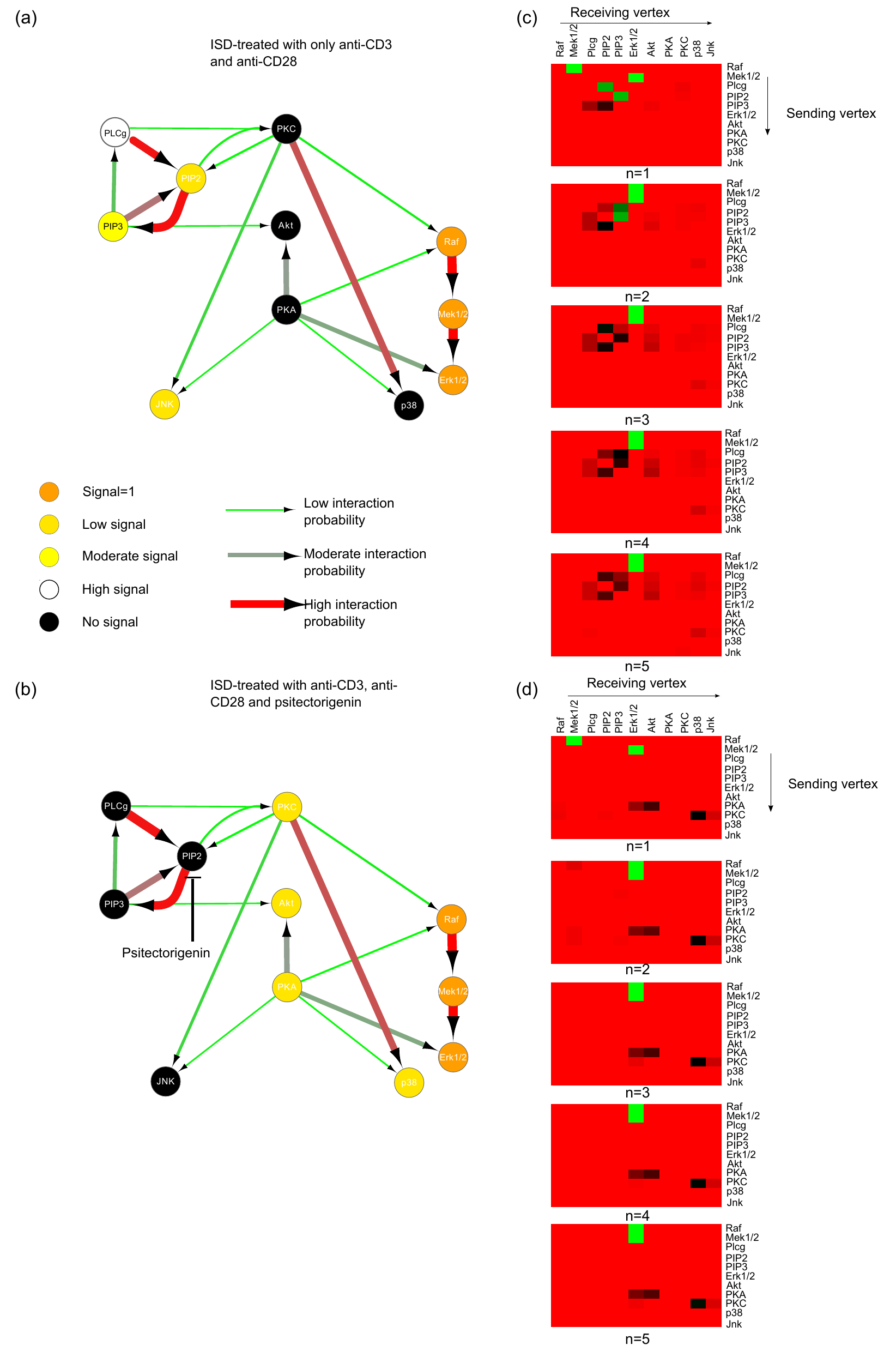}
\caption{(Color Online): The left hand side of the figure shows the ISD and edge weights for the anti-CD3 and anti-CD28 treated network (a) and for the anti-CD3, anti-CD28 and psitectorigenin treated network (b). The right hand side displays matrices of NTEs computed between every vertex pair over path lengths $n=1-5$, in the anti-CD3 and anti CD28 treated network (c) and the anti-CD3, anti-CD28 and psitectorigenin treated network (d)}
\end{figure*}
\end{document}